\documentclass[conference,a4paper,twocolumn]{IEEEtran}

\IEEEoverridecommandlockouts

\ifCLASSINFOpdf
\else
\fi
  \usepackage{tipa}
  \usepackage[pdftex]{graphicx}
	\usepackage{epsfig}
	\usepackage{epstopdf}

\usepackage[cmex10]{amsmath}
\usepackage{subfigure}
\begin{document}



\title{An Enhanced Interleaving Frame Loss Concealment Method for Voice Over IP Network Services}


\author{\IEEEauthorblockN{
 Tarek Gueham and Fatiha Merazka
}
\IEEEauthorblockA{
LISIC Laboratory, Telecommunications Department\\
University USTHB\\
Algiers, Algeria\\
Email: gueham.tarek@gmail.com, fmerazka@usthb.dz  
}
}


%


\maketitle


\begin{abstract}
This paper focuses on AMR WB G.722.2 speech codec, and discusses the unused bandwidth resources of the senders by using a Word16(16 bit) to encode the sent frames. A packet loss concealment (PLC) method for G.722.2 speech codec is proposed in order to overcome this problem and increases the efficiency of this codec by improving the quality of decoded speech under burst frame loss conditions over frame-switched networks. Objective and subjective experimental results confirm that our proposed algorithm could achieve better speech quality. Our proposed method achieves a PESQ value higher than 2 at 20\% frame erasure rate and ensure the compatibility between our modified decoder and the non-modified G.722.2 coder.\end{abstract}



%
\IEEEpeerreviewmaketitle


\section{Introduction}
Internet Protocol (IP) network has become a universal communication network that will accommodate all types of traffic, including images, voice, video, and data. An elementary and challenging component among them is the transmission of voice frames. The transmission of frame voice over IP (VoIP) networks, is gaining much attention as a future alternative to conventional public switched telephone networks (PSTN).  The ability to reduce the cost of long-distance telephone calls and provide additional capabilities is attracting customers to this tool. 
However, coded speech frames are transmitted through the IP network on the best-effort basis, without any guarantee of performance for real-time multimedia applications.  The three main problems occurring in real-time VoIP applications are \cite{1}:
1) End-to-end delay: is the total delay experienced by the frame from the sender till it reaches the receiver. For example, if the frame network is congested frame delivery is delayed beyond predefined thresholds and thus resulting in frame arriving late.
2) Jitter: Refers to the variation in frame inter arrival time. This difference is between when the frame is expected and when it is actually received is jitter. 
3) Packet loss: Loss of voice frames from sender to receiver. 
The total packet loss can be seen either  packets lost over the network due to congestion or packets arriving late after their expected playout time that are discarded by the receiver. The jitter caused by variable delays in the network is ultimately translated into the effect of frame loss in the network, as the frames arriving after the playout time are considered as lost. 
Research on the quality of audio transmission has focused on designing frame loss concealment (FLC) system which dynamically adapts the behavior of the audio application to maximize the audio quality under the constraints of restricted bandwidth, frame loss, delay and jitter present in the network. 
The FLC system can consist of varying mixtures of sender-based and receiver-based strategies. Sender-based schemes require the contribution of the transmitter, while receiver-based schemes are limited to the receiver. Sender-based schemes are mainly based on transmission of redundant information such as the forward error correction \cite{2}, which is an attractive way to increase the reliability of speech frames and to reduce the necessary time to recover the lost frame. Another common approach is to transmit two descriptions representing the same speech signal portion, where the first one is a high quality description and the second redundant has a low bit-rate description onto one of the neighboring frames \cite{3}. Multiple description coding can be seen as a bandwidth efficient generalization of these techniques \cite{4}.
Other researchers propose to add some extra information in the next frame such as linear prediction coding (LPC) coefficients \cite{5},energy and zero crossing ratio \cite{6}, voicing and fundamental frequency information (F0) \cite{7}, excitation parameters added to the preceding frame \cite{8} to protect only important frames. The use of additional side information as a means for improving concealment performance has been demonstrated for code excited linear prediction (CELP)-based coders such as G.729 \cite{9} and AMR-WB \cite{10}.
Receiver-based FLC techniques attempt to recover the speech signal content of a lost frame from its neighbors.
A common and simple method to recover lost frames is inserting a stand-in frame. This stand-in could be: a silence frame or a noise frame, or the repetition of the last received frame. The simplicity and the low latency requirement are a big advantage for this method, but the inevitable artifacts and the sudden noticeable transition between natural and synthesized speech introduced by those methods make the perceptual quality of speech is not significantly improved \cite{11}. Another approach is to use interpolation techniques where the parameters of neighboring frames are exploited to produce the replacement for the lost frame \cite{12}. Such methods are based on a statistical interpolation (SI) algorithm \cite{13}, which applies an interpolation in the compressed modified discrete cosine transform (MDCT) domain, by treating each time-trajectory of the coefficients for a given frequency bin as a separate signal with missing samples. Or, the discrete short time Fourier transform (DSTFT) domain \cite{14}\cite{15}  those methods use a complex spectral domain, where the signal representation is less fluctuating, whereas in the MDCT domain, the coefficients typically show rapid sign changes from frame to frame in each frequency bin \cite{16}. Interpolation in the DSTFT domain requires the conversion from MDCT to DSTFT. Such conversions add complexity to the decoder, and even though efficient conversions were developed and used in \cite{14}, the overall complexity is still quite high in addition to the low efficiency against consecutive losses. Unlike the above cited methods, time scale modification is a packet loss concealment technique based on time domain processing; the packets above the missing packet are extended without modifying the pitch frequency of speech signal. Almost all the TSM methods use the overlap and add (OLA) algorithm that does not analyze the content of the input signal just overlap and add the signal. Synchronous overlap and add (SOLA) algorithm is the enhanced version of the OLA algorithm. But SOLA does not maintain maximum local similarity. Waveform similarity and overlap add (WSOLA) is the technique that ensures sufficient signal continuity at segment joins that existed in original signal \cite{17}\cite{18}.
Waveform substitution is another approach to overcome the packet loss problem; those methods are based on speech signal stationary characteristic. It uses the frames prior to the lost frames and tries to use the most recent ones. It examines buffered frames and searches for the best match \cite{19}. 
Another interesting technique is to exploit both, past and future frames present in the jitter buffer such as using hidden Markov models (HMM) for estimating lost frame parameters is presented in \cite{20}; or using audio inpainting to reconstruct missing parts of audio signal basing on the good received parts, in condition that the duration of missing parts is lower than 50ms \cite{21}. Sparsity based techniques \cite{22} or, based on self-content \cite{23}. In this paper we propose an algorithm that exploits the bit representation of G.722.2 coded frames to insert a part of the previous frames to recover them in case of their lost.\\
We structure the remainder of this paper as follows. We begin by describing the bit representation in the G.722.2 codec and outline the limitation provided by this representation. Then, we introduce our proposed frame loss concealment method. Then, we show how this improves the robustness of the G.722.2 codec to frame losses, where the wideband Perceptual Evaluation of Speech Quality Mean Opinion Score (WB-PESQ MOS),  Enhanced Modified Bark Spectral Distance (EMBSD), and MUltiple Stimuli with Hidden Reference and Anchor (MUSHRA)  are used as the subjective and objective quality metrics for performance evaluation. After that, we compare our proposed method with the forward error correction (FEC) scheme to evaluate its performance. Finally we conclude.

 


\section{Bit representation in G.722.2 codec}
With the wider passband of 50-7000 Hz, wideband speech provides much better quality and naturalness compared to narrowband telephone bandwidth speech. The G.722 codec encodes wideband speech by first decomposing it into two subbands, and then using backward adaptive differential pulse code modulation (ADPCM) to code each subband. At the receiver, the coded speech in each subband is decoded using an ADPCM decoder, and reconstructed to give the decoded wideband speech \cite{24}. This simple coding and decoding operation causes an algorithmic delay of only 3 ms and provides good performance at rates of 48, 56 and 64 kbps \cite{25}\cite{26}. The low delay, low complexity and high quality of the G.722 wideband speech codec have resulted in its adoption by several VoIP and Voice over Wireless LAN (VoWLAN) phones \cite{27}\cite{28}.

\begin{figure}[!ht]
\centering
\includegraphics[scale=0.6]{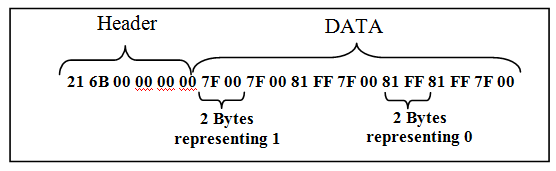}
\caption{The first 26 bytes of a G.722.2 frame}
\label{fig01}
\end{figure}

We notice that the G.722.2 codec uses a Word16 (16 bit) to code a single bit of the frame (7F 00 to represent 1 and 81 FF to represent 0). For example, the G.722.2 codec uses 135 bits to code a single frame. But in reality, it uses 270 bytes to code it \cite{29}. And we consider that, this is a big waist of sender resource.  
Fig.~\ref{fig01} represents the first 20 bytes of a frame encoded with G.722.2 codec mode 0 and represented in a hexadecimal format where each character represents 4 bits; the first 6 bytes represent the frame’s header. Then, we notice that the bit 1 is represented by “7F00” and the bit 0 is represented by “81FF”.

\section{The proposed algorithm
}
\label{sec:2}
\begin{figure}[!ht]
\centering
\includegraphics[scale=0.5]{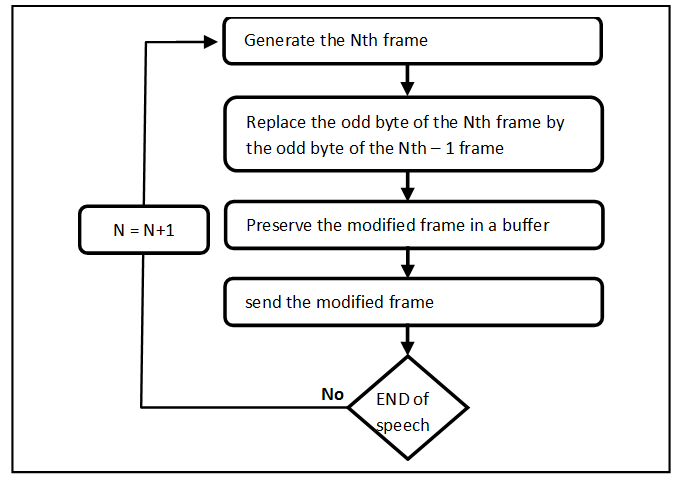}
\caption{Algorithm at Sender side}
\label{fig02}
\end{figure}

\begin{figure}[!ht]
\centering
\includegraphics[scale=0.7]{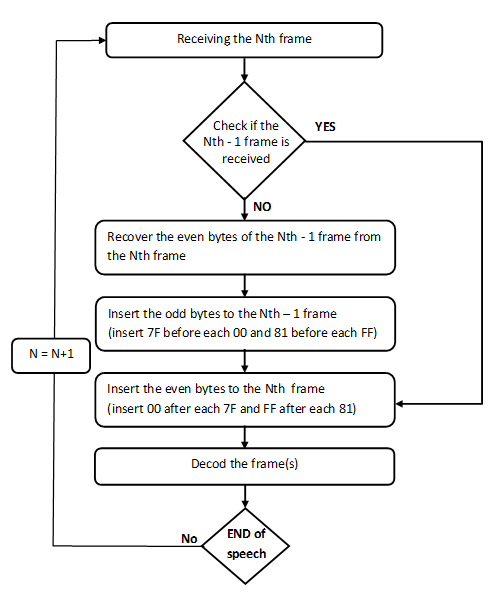}
\caption{Algorithm at Receiver side}
\label{fig03}
\end{figure}

In order to exploit effectively the resource of the sender and protect the sent frames against loss; we propose to replace the even bytes of the current frame by the even bytes of the previous frame, those even bytes will be used to reconstruct the previous frame in case of their lost. The proposed method will be detailed in Figs.~\ref{fig02} and~\ref{fig03}.
\subsection{At the sender}
\label{sec:1}
After the generation of the Nth frame, the coder has to replace every even byte (that has the value 00 or FF) of this frame by the even byte of the Nth-1 frame (that has the value 7F or 81), which means:
\begin{equation}
frame (N ,j) = frame (N-1,j+1)
\label{eq01}
\end{equation}
where:
\begin{itemize}
\item $frame$ : represents a G.722.2 frame.
\item $N$ : represents the number of the last generated frame.
\item $j$ : represents the byte number of the Nth frame, and $j = 8, 10, 12, …, 270$ in mode 0. 
\end{itemize}

After that, the coder has to put the Nth-1 frame in a FIFO buffer of one frame size (270 bytes in G.722.2 codec mode 0).

\subsection{At the receiver}
\label{sec:1}
After receiving the Nth frame, we check if the Nth-1 frame has been received. 
If not, we recover the Nth-1 frame, by inserting the even bytes of the Nth frame to the even bytes of the Nth-1 frame. This means:
\begin{equation}
frame (N-1 ,j) = frame (N,j)
\label{eq02}
\end{equation}
Then, we recover the odd bytes of the Nth-1 frame by inserting the values “7F” or “81”:
\begin{equation}
if  frame (N-1 ,j)= FF:  frame (N-1 ,j-1) = 81 
\label{eq03}
\end{equation}
\begin{equation}
if  frame (N-1 ,j)=00:  frame (N-1 ,j-1) = 7F
\label{eq04}
\end{equation}
After recovering the Nth-1 frame, we have to normalize the Nth frame to avoid any errors by inserting the values “00” or “FF” in the even bytes:
\begin{equation}
if  frame (N-1 ,j)= FF:  frame (N-1 ,j-1) = 81
\label{eq05}
\end{equation}
\begin{equation}
if  frame (N-1 ,j)=00:  frame (N-1 ,j-1) = 7F
\label{eq06}
\end{equation}
Now, if the Nth-1 frame is not lost, we execute just the previous step.
The advantage of this method is that we could protect the frames against loss without adding any extra information. Moreover, the modified G.722.2 decoder is compatible with  non modified G.722.2 coder and in this particular case the decoder will insert the Nth frame to conceal the loss of the Nth-1 frame (which is the common repetition method \cite{30}). 


\section{Simulation results}
\label{sec:3}
In this section, we assess the performance of our proposed method using WB-PESQ MOS,  EMBSD and MUSHRA  as the subjective and objective quality metrics. The experiments were performed using speech files of male and female speech as input. Each segment is comprised of two sentences by the same speaker and is 8 seconds in duration. The default operating rate of 6.60 kbps is used for the G.722.2 codec. Each frame comprises a G.722.2 coded speech frame of duration 20 ms and size of 264 bytes. The frame loss rates (FLRs) considered were 0\%, 2\%, 4\%, 6\%, 8\%, 10\%, 12\%, 14\%, 16\%, 18\% and 20\%. For each FLR, 10 runs of the experiment were performed to simulate different frame loss patterns in the speech files. The loss simulation were performed using Gilbert Extended Model (EGM), this extended model allows us to describe loss bursts of up to $m$ frames \cite{31}.
The average WB-PESQ calculated over the 10 different frame loss patterns corresponding to each FLR has been used as the quality measure in our experiments. The use of WB-PESQ is based on the results in \cite{32} that have verified the suitability of using WB-PESQ for evaluating wideband speech quality coded using G.722.2 for FLRs up to 10\%, and to assess the distortion between the original and the decoded signals we used the EMBSD as the distortion evaluation metric. Finally, for subjective test, we used the MUSHRA for listening quality evaluations. Ten listeners gave score according to quality of decoded by original and improved algorithm. The test sentences were presented to listeners at a randomized order.

\section{Comparing our proposed method with the forward error correction scheme
}
\label{sec:4}
\begin{figure}[!ht]
\centering
\includegraphics[scale=0.4]{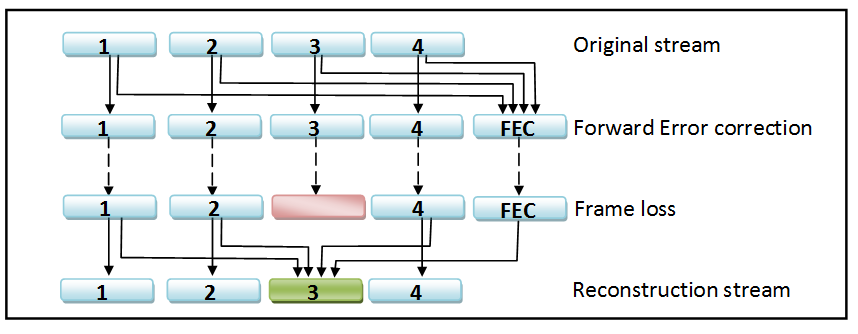}
\caption{FEC scheme}
\label{fig04}
\end{figure}

In order to evaluate our proposed method, we compared it with firstly, the original G.722.2 codec, this comparison is fair because our proposed method doesn't modify the bit rate of the sender neither the number of sent packets. Secondly, 
we compared it with an existing method which is the FEC parity coding scheme. 
A number of FEC techniques have been developed to conceal losses of data during transmission \cite{33}. These schemes rely on the addition of conceal data to a stream, from which the contents of lost frames may be recovered. In parity coding, the exclusive-or (XOR) operation is applied across groups of frames to generate corresponding parity frames. An example of this has been implemented by Rosenberg \cite{34}. In this scheme, one parity frame is transmitted after every 4 data frames. Provided there is just one loss in every 5 frames, that loss is recoverable. This is illustrated in Fig.~\ref{fig04}.

\begin{figure}[!ht]
\centering
\includegraphics[scale=0.4]{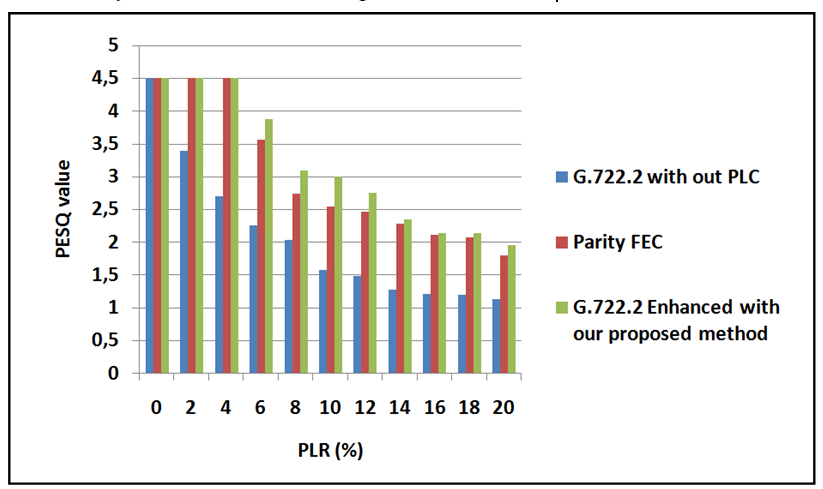}
\caption{Comparison between our proposed method and the FEC parity scheme using PESQ metric}
\label{fig05}
\end{figure}

Fig.~\ref{fig05} shows a comparison between our proposed method and the Parity FEC scheme using PESQ metric.

\begin{figure}[!ht]
\centering
\includegraphics[scale=0.4]{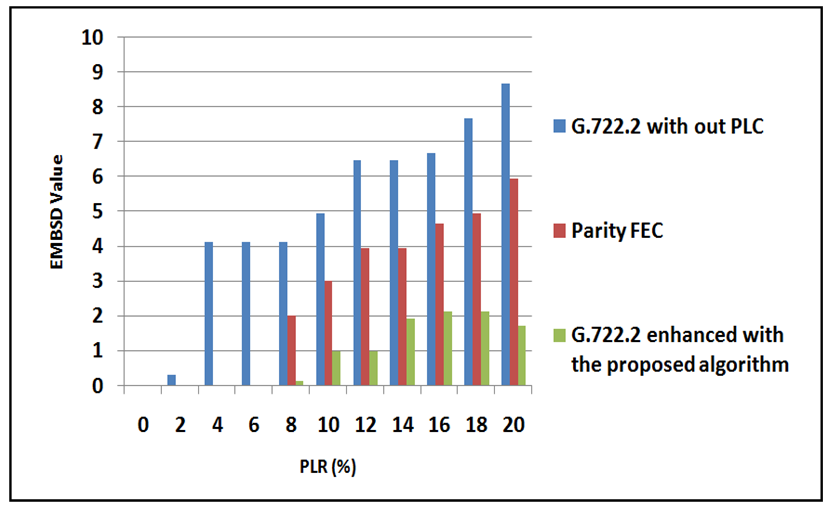}
\caption{Comparison between our proposed method and the FEC parity scheme using EMBSD metric}
\label{fig06}
\end{figure}

\begin{figure}[!ht]
\centering
\includegraphics[scale=0.37]{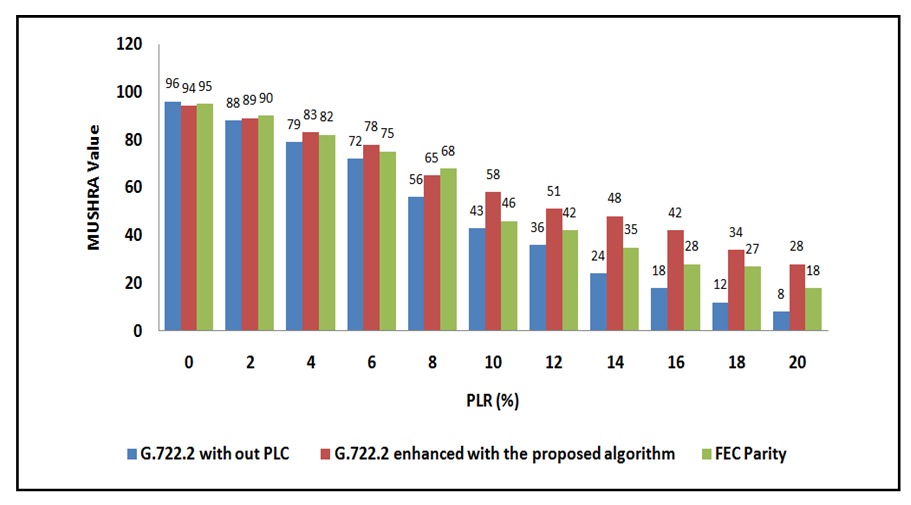}
\caption{Comparison between our proposed method and the FEC parity scheme using MUSHRA metric}
\label{fig07}
\end{figure}

Fig.~\ref{fig06} shows a comparison between our proposed method and the  FEC parity scheme using EMBSD metric.
Fig.~\ref{fig07} shows a comparison between our proposed method and the FEC parity scheme using MUSHRA metric.

The metrics values are slightly higher for our proposed method than the FEC parity due to the limitation of FEC schemes where consecutive frames are lost and where just one loss in every 5 consecutive frames is allowed to achieve the recovery. On the other hand, the FEC parity scheme implement an additive redundancy traffic (20\% of the original traffic), plus the additive delay needed to recover the lost frames, unlike our proposed method which wait the reception of the subsequent frame to recover the lost frame, all that without adding any redundancy traffic. The disadvantage of our proposed method is that it recovers just the last lost frame in a sequence of frame loss.

\section{Conclusion}
We have proposed and investigated the performance of a new FLC scheme for the standard G.722.2  speech codec. While the experiments have been performed on the G.722.2 speech codec mode 0, the proposed scheme is clearly applicable to other modes or codecs as well. Our results show a significant improvement in the performance of the G.722.2 speech codec. The proposed algorithm achieves a PESQ value higher than 2 at 20\% frame erasure rate and ensure the compatibility between our modified decoder and the non-modified G.722.2 coder.




%


\end{document}